# Anticipation-driven Adaptive Architecture for Assisted Living

**Mihai Nadin[1*], Asma Naz[2]**
Institute for Research in Anticipatory Systems, University of Texas at Dallas, Richardson, Texas, USA
[2]Department of Architecture, Bangladesh University of Engineering & Technology, Dhaka, Bangladesh
*** Correspondence:**
Dr. Mihai Nadin
nadin@utdallas.edu

**Abstract:** Anticipatory expression underlies human performance. Medical conditions and, especially, aging result in diminished anticipatory action. In order to mitigate the loss, means for engaging still available resources (capabilities) can be provided. In particular, anticipation-driven adaptive environments could be beneficial in medical care, as well as in assisted living for those seeking such assistance. These adaptive environments are conceived to be individualized and individualizable, in order to stimulate independent action instead of creating dependencies.

## 1 Goal

As a proof-of-concept contribution, the following are the specific aspects to be addressed:

A. What makes anticipation-driven adaptive architecture (ADAA) necessary.
B. What makes ADAA possible and different from current attempts to provide living spaces for medical care and for assisted living.
C. How the goals of ADAA are achieved (concept details, functionality, feasibility).
D. Design of an ADAA working prototype.

Each will be specifically addressed (see 1.1, 1.2, 1.3, 1.4).

### 1.1 *The IS situation*

The numbers are worrisome

- Cognitive deterioration: About 70% of the population ends up in need of assisted living, largely due to declining cognitive ability (Tajana Simunic Rosing, 6th IBM Research Cognitive Colloquium, September 18-19, 2017, Yorktown Heights, NY). The Center for Disease Control and Prevention (CDC) declared cognitive decline a public health issue.
- Rapid aging demographics: The percentage of the population 60 years and above will double (from ca. 605 million to 2 billion) in the next 20 years. A comprehensive source of data helpful in defining the current situation is *World Population Aging 2019* from the United Nations Department of Economic and Social Affairs Population Division.
- Available resources: Healthcare delivery and assisted living require qualified personnel. Not enough physicians and caregivers are trained for this condition. America will face a shortage of up to 122,000 physicians by 2032 and will need to hire at least 200,000 nurses per year to meet increased demand and to replace retiring nurses (AHA Fact Sheet: Strengthening the Health Care Workforce, May 2021). The predicable shortage affects affordability and suggests the need for alternatives. The same situation is reported in Europe. The extreme shortage of medical personnel in Africa and in large parts of Asia will become even more dire.

### 1.2 *Anticipatory action*

Improved knowledge of anticipatory processes is increasingly becoming available (Nadin 2019), (Nadin 2021). In particular, the quantification of anticipatory expression using the *AnticipationScope* resulted in a repository of data that can inform a variety of new applications. Knowledge regarding adaptive spaces created a premise for designing such spaces (Naz et al. 2017) (Naz 2017) (Nadin, Naz 2018) (Nadin, Naz 2019). Moreover, there is progress in the area of intelligent materials. Suffice it to mention that *The Journal of Intelligent Materials Systems and Structure* (an international peer-reviewed journal reporting the results of original experimental and theoretical work on any aspect of intelligent materials systems) documents this progress. Smart structures, smart materials, active materials, adaptive structures, and adaptive materials are more and more deployed. The intelligent space discussed herein can be adapted via IoT technology at various scales: a whole wall or only part of it, a whole room, a whole facility, etc.).

The major difference between ADAA and the most successful applications in use is the holistic nature vs the fragmentary (part separated from the whole) aspect. Indeed, health monitoring, smart devices, and virtual companions are viable partial solutions. Quite often, they do not stimulatemore independence, but rather create dependencies. Ambient Assisted Living (AAL) (Monekoso et al. 2015, Abrantes 2021) provides means for monitoring those in need, but not for adaptive accommodation.

The non-invasive living environment (or medical facility, e.g., emergency room, surgery suite, etc.) that ADAA makes possible is not reactive, as almost all AI-based solutions are, but anticipatory. It provides individualized living assistance or recovery (after medical intervention) in the form of an intelligent environment. In this environment, assistance is a service provided in anticipation of actions, not in reaction to situations as they arise. For example: Prior to rest periods, the system recognizes the preparation: lights are dimmed, sound volume is reduced. Before surgery, a calming environment (e.g., room color, lighting, calming sounds) contributes to preparation. During surgery, the whole environment (space, medical team, equipment) is taken into account in order to adapt to a variety of possible situations.

The major challenge is the ability of the system to interpret anticipatory cues. For example, in preparing for a shower, the person gets undressed. But there are other situations in which a similar act might occur. Therefore, the Anticipatory Profile helps the system to deal with ambiguity.

### 1.3 *Integrating anticipation and machine learning*

ADAA is premised on the Anticipatory Profile (AP) of each individual. Everything a person does, from the simplest movement (e.g., lifting a hand) to more elaborate actions (e.g., taking a shower, cooking, driving) engages the entire organism. The Anticipatory Profile is the mapping from a person's cognitive, physiological, and motoric system to digital representation associated with a Machine Learning (ML) facility. An AP is established once an individual enters into an assistive condition. The progressive change in the AP is reflected in the patterns of behavior that translate into data from ADAA to the ML module. Each “thing” in the ADAA (water faucets, electric switches, medical monitoring devices, chairs, beds, etc. etc.) is part of the IoT ledger subject to control processes guided by the ML module (Figure 1).

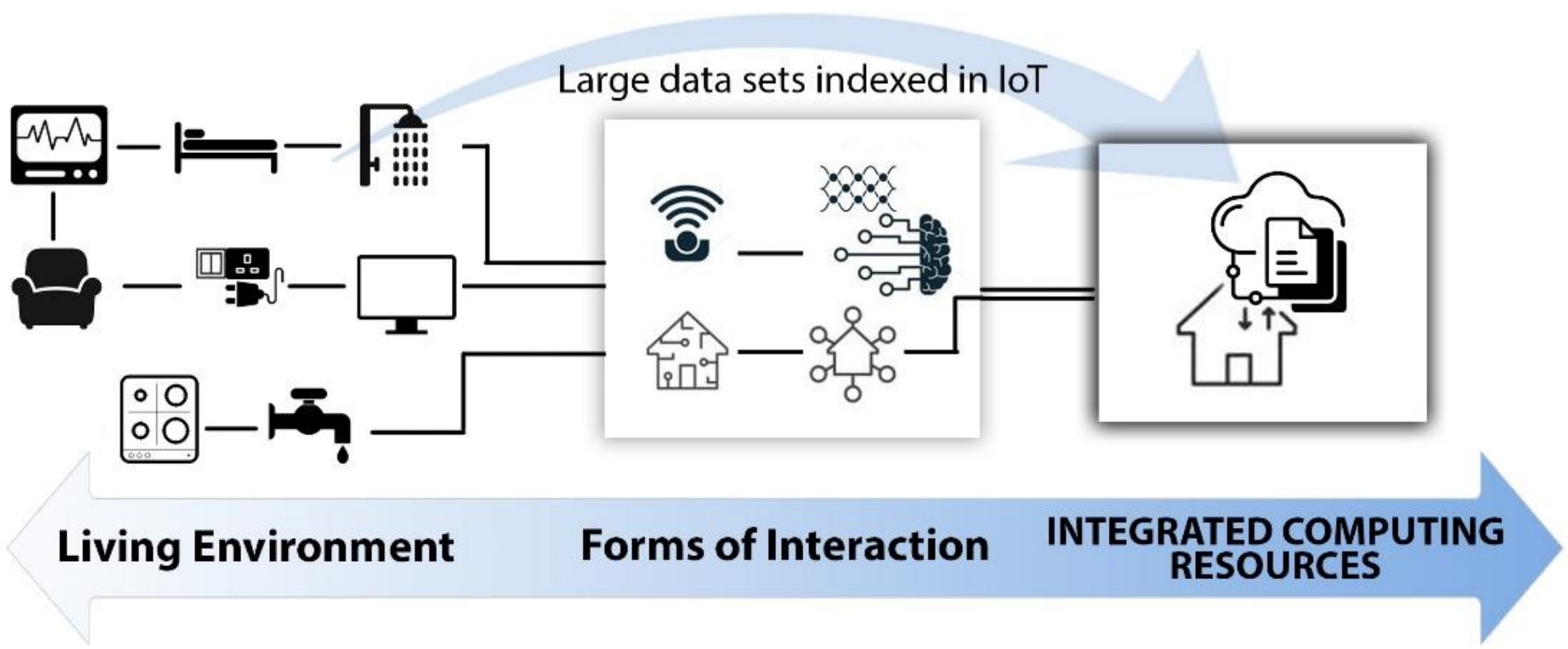

Figure 1. Possible activities (resting, watching TV, reading, washing) are referenced in the large data set generated through sensor acquisition. They are indexed in IoT (ledger).

### 1.4 *The narrative of anticipatory assistance*

Take these examples: Preparing to go to sleep entails a change in the “feeling” of a space—sound level, lighting ventilation. In preparation for a shower or bath, several preliminaries may be considered: getting weighed, toilet use, room temperature, water temperature, the possibility of an accident. Communication with professional caretakers (virtual or otherwise), with family members, and with friends is supported through a variety of media (sound, sound and image, etc.). The system keeps a record to be used for training purposes. In the living space, a virtual window, with images of emotional significance, can be conjured. A virtual visit with family members or friends can take place using the currently deployed applications that have been tested in retirement facilities (under the pressure resulting from Covid-19). The entire project translates the model of anticipation (Figure 2) into an integrated large data processing “facility” that has as output the adaptive space and the assisted living performance (Figure 3). This is an integrated intelligent virtual reality that uses intelligent materials instead of goggles.

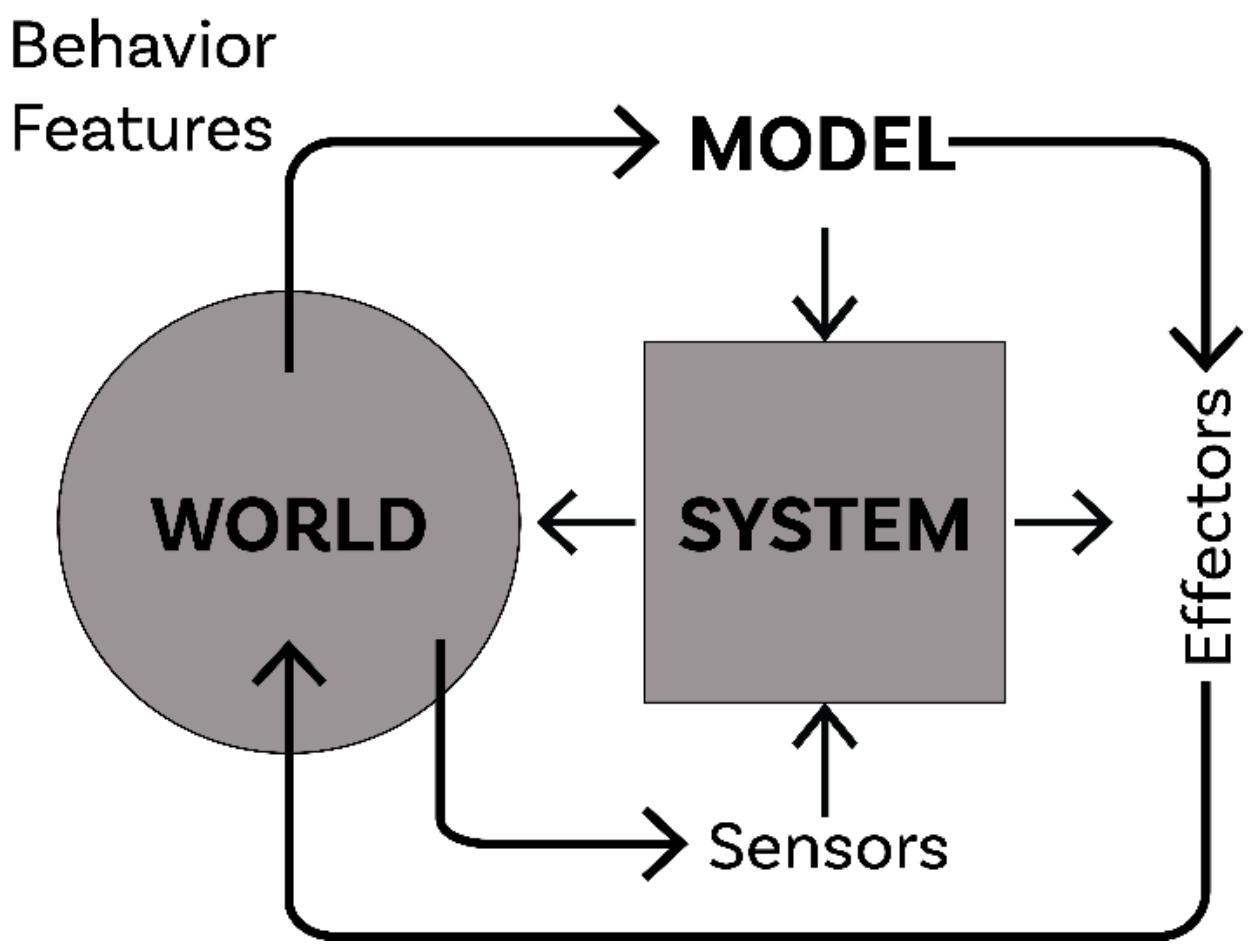

Figure 2. Implementing an anticipatory system.

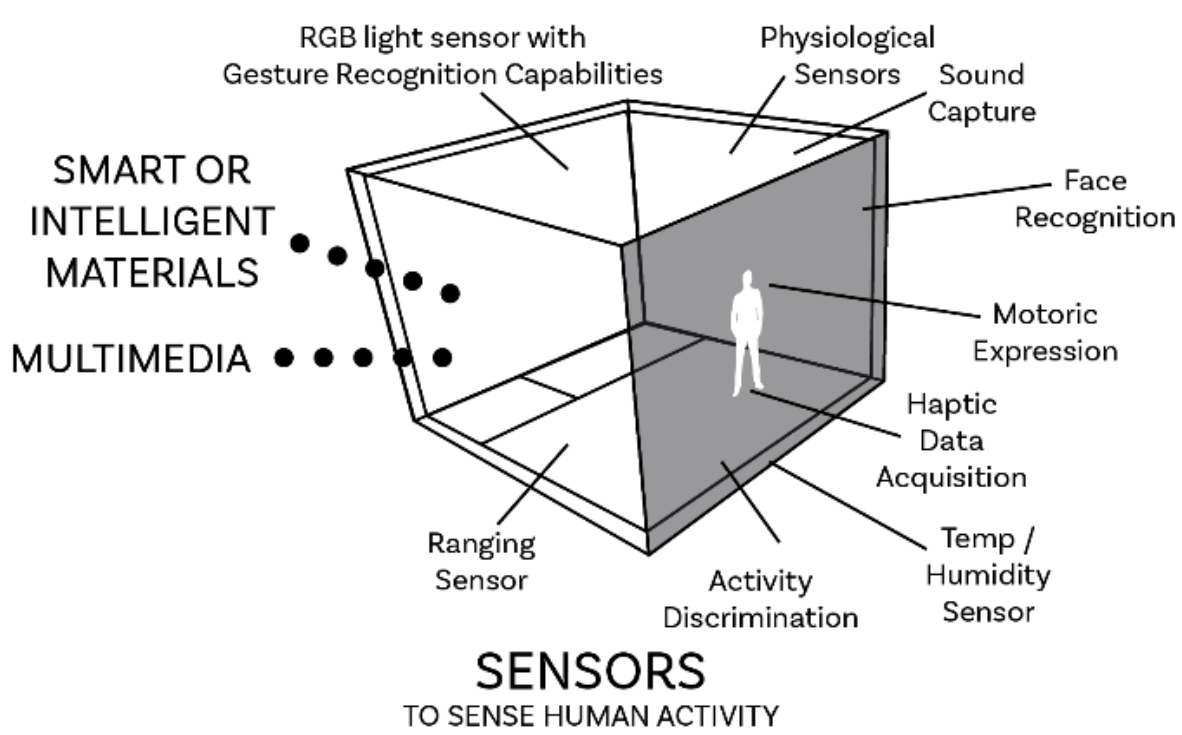

Figure 3. Structural representation of the adaptive space for assisted living. An anticipatory system contains a model of itself that unfolds in faster than real time. Sensors and actuators allow for data acquisition and for consequent actions on the appropriate object (e.g., the faucet, for example, before and after it was used) in the adaptive space.

## 2 Preliminary work

Losses in sensory acuity and discrimination, motor control, cognitive functions, memory performance—all contribute to diminished anticipation capabilities. Therefore, ADAA aims to compensate for these by stimulating brain plasticity.

Since 2004, *Project Seneludens*[1], engaging the aging through “playful” activities (under IRB approval), data from 233 subjects have been acquired in order to develop an understanding of the various aspects of aging and the types of assistance it requires. The gist of *Seneludens* is to compensate for aging-related losses in memory, motor control, sensation, and cognitive abilities, by developing the human being's innate brain plasticity. This could be attained through targeted behavioral training in rich learning environments. In particular, specially conceived, individualized computer-facilitated interactions that address *both* cognitive and motoric abilities can help in attaining this goal. For the subsequent ADAA project, the knowledge accumulated in describing how actions are performed in order to provide an adaptive assistive environment will be accessed and applied.

[1] https://seneludens.utdallas.edu

## 2.1 *The AnticipationScope*

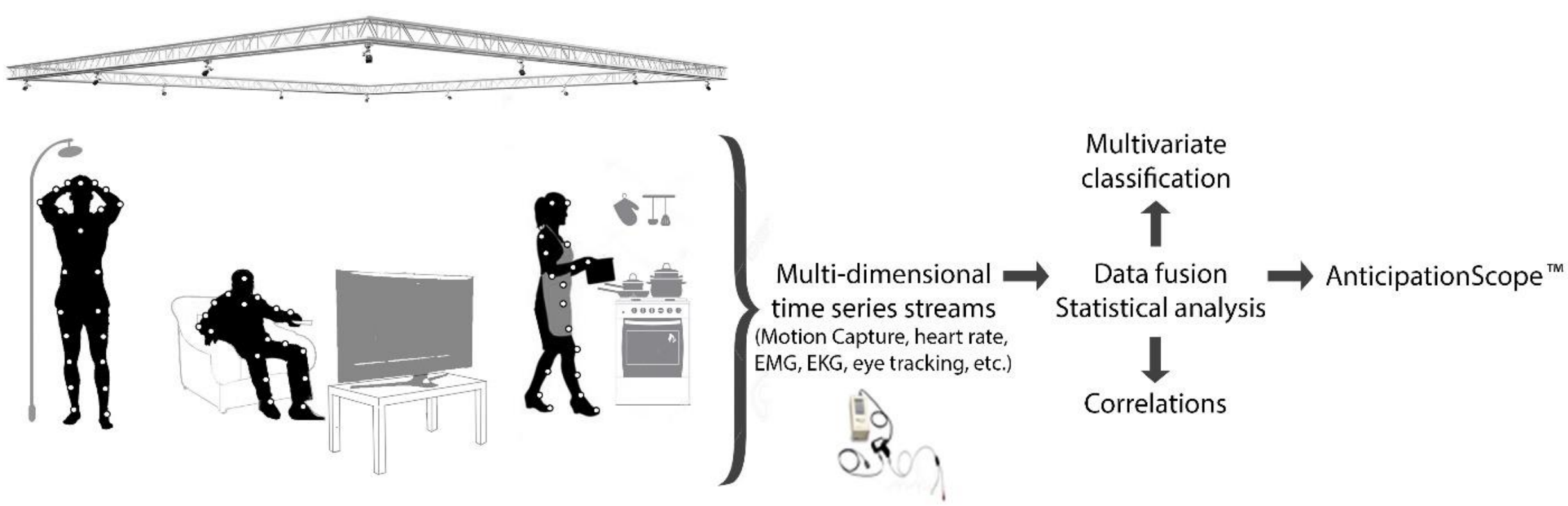

Figure 4. The *AnticipationScope* for data acquisition and for testing.

The AnticipationScope (Figure 4) produces a representation of the human being in action, as an integrated expression across multiple systems, including sensory perception, proprioperception, cognition, memory, motor control, and affect. The AnticipationScope can be corelated to a brain imagery device: data collected through the AnticipationScope can be correlated to that gained through the "brainscope."

The AnticipationScope quantifies the process by capturing the motion, not on film or video, but in the mathematical description (as a matrix) corresponding to distributed high-resolution motion capture technology. Such a description allows focusing on single points on the body, and to pick up even the most evasive tremor or motion characteristics (of joints, body segments, muscles, etc.). Sensors, such as EMG, goniometry, accelerometry, blood pressure, EEG, etc., were applied to different parts of the body. Such sensors capture the various *preparatory* processes, as well as the reactive components, of maintaining balance or of initiating new actions (preparing for rest, for a shower, performing various tasks related to independent living).

To describe how anticipatory characteristics, which underlie our adaptive capabilities, change over time implies the need to develop means not yet available for quantifying anticipation. With such means, the many inter-related factors involved in adaptive deficiency can be identified, and eventually retraced to the molecular, neuronal, or DNA level. None of the conditions mentioned results from diminished reaction. All are expressions of affected, mainly reduced, anticipation. Since anticipation is an integrated expression of each individual's characteristics across multiple systems, the goal was to conceive, design, specify, test, and implement an anticipation scope— the metaphoric equivalent of a microscope—that can help "see" how anticipation takes place. I conceived of the AnticipationScope as an integrated information acquisition, processing, interpretation, and clinical evaluation unit (Nadin 2013).

### 2.2 *Perceptual studies in the virtual environment*

Independent of the utilization of the AnticipationScope, a user study in a six-sided projected immersive display was conducted to examine emotional responses to perceived spatial experiences (Naz et al. 2017). Quantitative and qualitative correlations were established based on the variable design parameters (light, color and texture) and perception (Figure 5).

Thirty-two participants rated each space on a scale of 1 to 10 (1 the lowest, 10 the highest) in terms of degree of psychophysiological aspects of space: warmth, coolness, spaciousness, intimacy, excitement, calmness, comfort, preference for rest and work. Students and faculty from different disciplines, as well as staff, in a major university participated in the project. Twenty-nine participants (91%) ranged in age from 18-40 years (Mean=27, SD=4.8); three (9%) were in the age range of 50-76. A single room was simulated in a mixed reality environment. The design parameters had three attributes: color (orange and blue), brightness (dark and light), and texture (rough and smooth). Eight virtual spaces—composed of one characteristic of each design attribute—were presented to each participant in a pre-defined random order. Since perception depends on context, half the number of the participants were told to sit still during the test; the other half carried out some activity (e.g., folding clothes) during test.

A mixed-design factorial (ANOVA), with three repeated-measures factors and one between-subjects factor, was carried out. Data were reported as statistically significant at $p<.050$. Data analysis also revealed that the subjective response to the virtual space was consistent with corresponding responses from real-world color and brightness perception (Franz 2006). As in real-world perception, orange (warm colors) was found warmer, more exciting and intimate than blue (cool colors), which was found cooler and calmer. Also similar to real-world perception, brightness had a significant impact on perception of spaciousness.

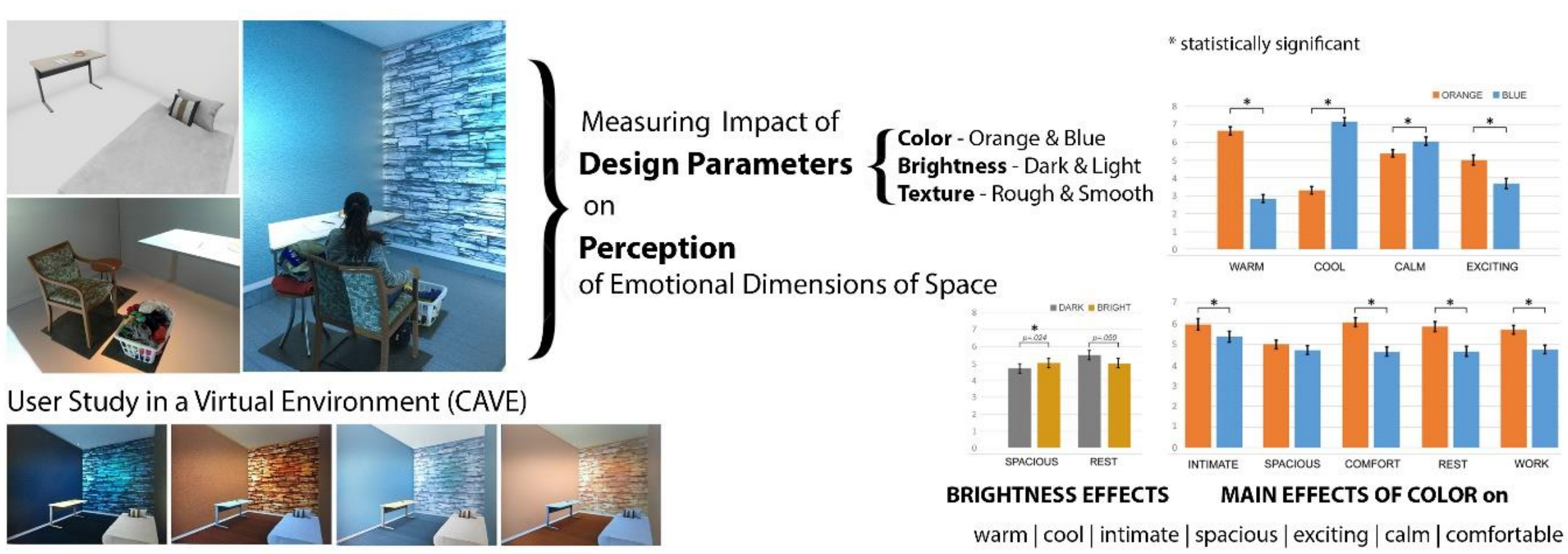

Figure 5. Perceptual data acquisition.

### 2.3 *Machine Learning perception-based emotional space design*

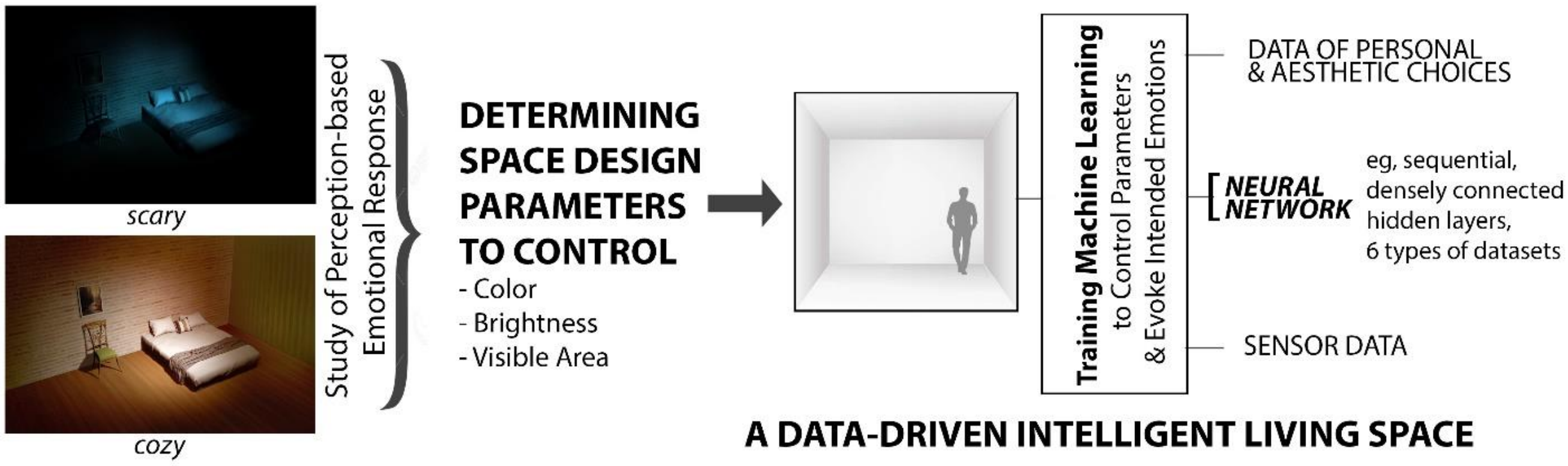

Figure 6. Training Machine Learning with the purpose of adaptive performance.

The following examination (Figure 6) was conducted on training Machine Learning (ML) as an interface for designing perception-based emotional spaces (Naz 2021). It was conducted in two parts. The aim of the first part of the study (Study-I) was to identify the extent to which certain space design parameters can be manipulated in order to evoke specific emotions. The second part of the study (Study-II) used this data to train ML in creating specific environments by manipulating those design parameters

For Study-I, two computer-rendered still images of the same scene were created by manipulating the properties of light (color, brightness and visible area) in order to evoke in the observers two distinct and opposite emotional responses: "scary" and "cozy." This was a within-subject, qualitative, online user study with 22 participants (14 males and 8 females, age range = 25 to 50 years, M=34.50, SD=8.17). The findings revealed that twenty-one participants indeed found the spaces evoke the expected feelings.

Study-II used the data (the exact values for color, brightness and visible area) and props from Study-I to simulate a scene in "Unity." A user test was run via server in order to collect a total of 1400 datasets as the basic training data. Participants rated scenes according to degrees of "scariness" or "coziness." Measures were taken to avoid unnatural lighting. The Deep Learning (DL) model was designed on KERAS—an Open-Source Neural Network library written in Python—that runs on top of TensorFlow 2.0. This is a work in progress. The results of the first attempts looked promising. The degree of accuracy was low but the training loss, validation loss, and the predictions were satisfactory. A comparatively homogenous population (i.e., similar age range, demography, socio-cultural background, etc.) and more training data should improve accuracy.

## 2.4 *Addressing concrete goals for the medical community*

The data repository from the three data acquisition platforms used shows that most human actions are preceded by a preparatory phase. The individual is not aware of this phase, but it is essential to carrying through the act. ADAA is meant to pick up cues (pertinent to preparation) and engage each individual.

The research is focused on combined probabilistic and possibilistic inference based on sensor data pertinent to the adaptive patterns of individuals and groups. Combined probabilistic and possibilistic modelling, together with inference and learning in multi-scale and multi-purpose human action (specific to assisted living environments) make possible the generation of an adaptive profile. Generation of data in a wireless sensor network designed to capture cognitive, motoric and sensory characteristics, and harvesting of such data imply a procedure for data normalization and for data fusion. A distributed reasoner (intelligence at the sensor and actuator level) is performing the inference of the aggregate behavior.

Presentations were made for Children's Health Group and UTSW, December 14, 2018; Highland Homes (assistive living) Friday, January 25, 2019; McCarthy Construction (for hospitals and assisted living); and Dallas Health Group, January 30, 2019, followed by demonstrations of the technology.

The data acquired so far was used in a variety of applications. Assisted living could benefit from it. So could medical applications. This underlies the concept development undertaken in the last two years and presented in this study.

## 3 ADAA

The concept of affective space creation was initially introduced as an architectural inquiry into the specific needs of professionals, frequently on the move (also known as "neo-nomads") (Naz 2017). It is distinctly different from the reactive smart environments and addresses the anticipatory characteristics of ever-changing residence (Nadin 2012).

This communication presents a design concept on the basis of which an adaptive space was modeled. The space adapts to the changing needs of the life and work circumstances of patients undergoing medical treatment and for persons requiring assisted living or being assisted in cases of neuro-trama or in intensive care units. The word "adaptive" refers at this moment only to perception of spaces. An operating room and a medical recovery facility are supposed to accommodate different needs and expectations. Within the strict limitations of the physical size and shape, the adaptive space provides variable perceptual experiences for the user. These experiences result in perceived affective qualities of spaces (i.e., calm, soft, cozy, spacious, intimate, warm, cold, etc.) provided according to user needs and demands. The following example (Figure 7) demonstrates the modeling of variable perceptual experiences in a single room. The parameters acted upon are light, color, and texture. In the modeling process, design expertise provides choices tested with the user and subject to machine learning. Imagine an environment that feels "cozy," "bright," and "spacious" in which to perform some activity (e.g., checking the news, e-mailing). The same room may feel "calm" and "intimate" when it's time to rest, "contemplative" when reading a book, or "stimulating" when listening to music.

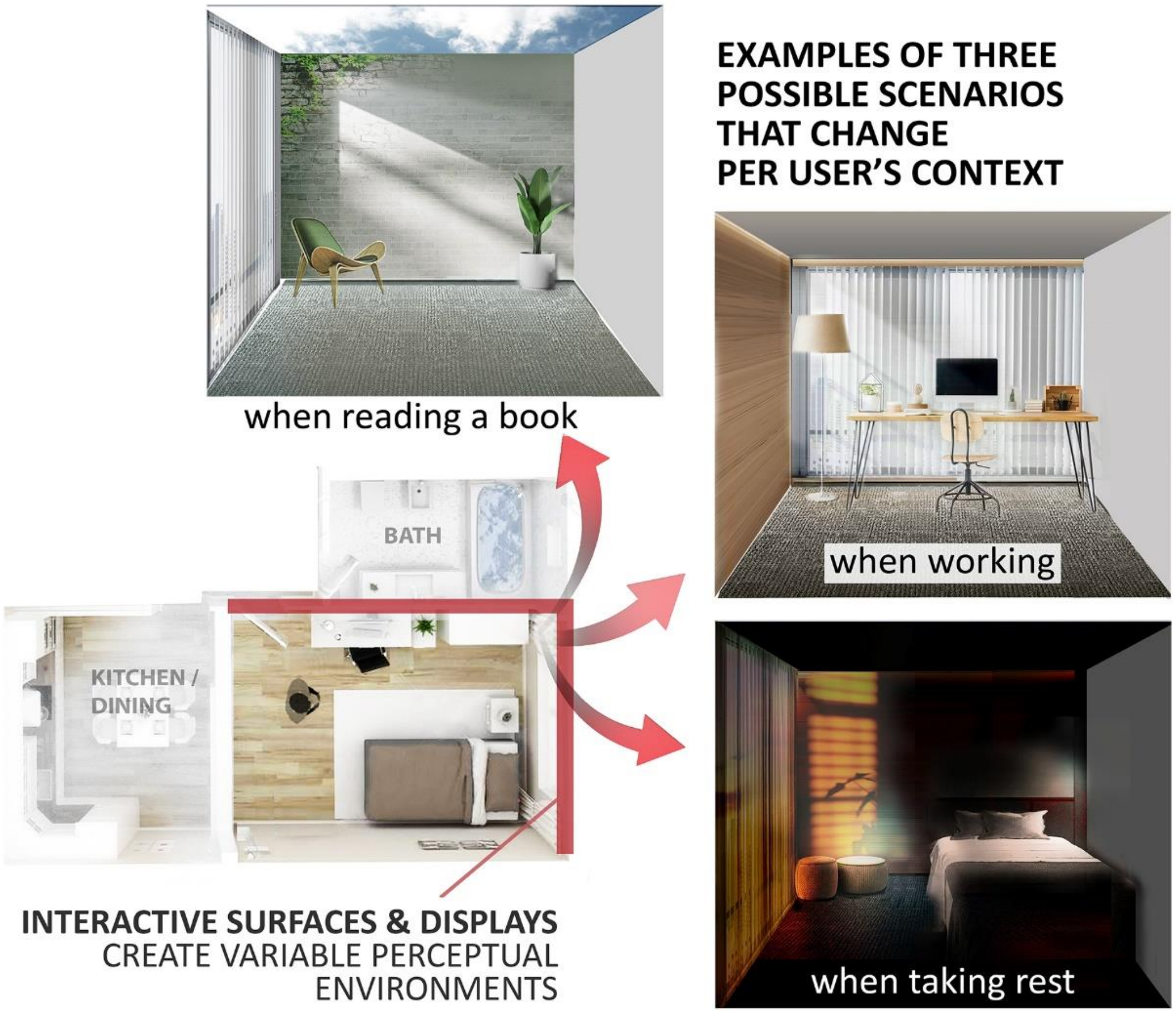

Figure 7: An anticipation-driven adaptive architecture.

ADAA generates such environments using only three enclosing surfaces: two adjacent walls and a ceiling. The affective or emotional dimensions are subjective and personalized while being data-driven. Due to the subjectivity and unpredictability of emotional responses, as well as an occupant's psychophysiological and functional needs, personalization will always evolve, influencing human behavior, actions, and creative goals (Nadin 2010). Intelligent materials with variable properties—temperature, reflection, refractions, hapticity—can interact intelligently according to context (Nadin, Naz 2019). For example, intelligent materials can synthesize "virtual" windows through video streams or as images. Using nanoparticles, CoeLux, a smart LED lighting system, simulates the sky and brings the effects of natural light inside interior spaces. The system can be programmed to imitate a wide range of "warm" and "cool" light and shadow conditions from different climatic zones around the world. Data drives this intelligent architectural solution (Figure 8).

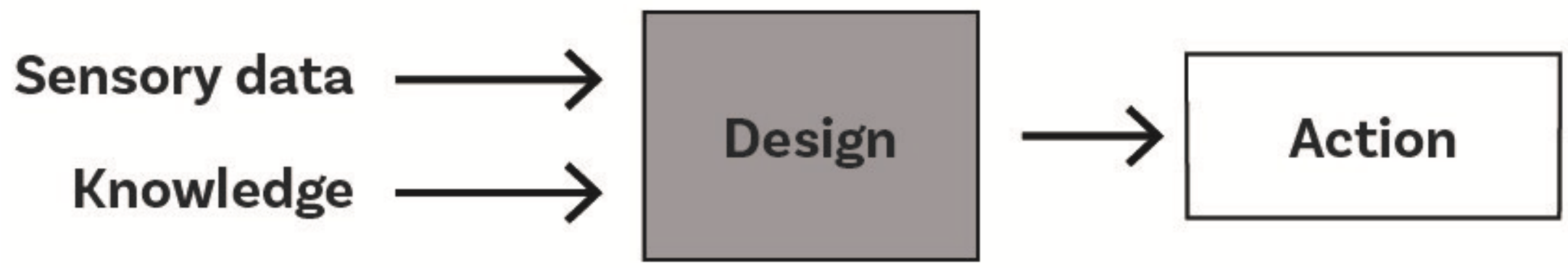

Figure 8. Sensory data and domain-specific knowledge (basically encoding principles of perception) drive the design machine which integrates machine learning. The output is the adaptive living environment (e.g., remote assistance in actions, change of space perception through production of virtual imagery via intelligent materials).

Data on user's aesthetic and personal taste, demography, socio-cultural status, medical records and activities, etc. inform the perceptual variability in the design. Hence, the space allows for individual expressions and personalization. The enclosing surfaces of the architectural space are endowed with interactive technologies—i.e., sensors for collecting data related to identifying emotions and activities, and effectors for creating variability of spatial experiences—integrated through an intelligent system. The design's sensorial capabilities collect any personal and environmental data relevant to the occupant's health, well-being, and activities, such as emotional state assessment, health monitoring, motoric expression, voice commands, activity recognition, temperature, humidity, dust sensors, haptic, physiological and gesture-recognition sensing, etc. With a network of ambient intelligence-based technologies (and cloud computing for computation resources and storage), the outputs of the effectors adapt through variable surface colors, brightness, visibility, and other displays. The Design Module is an "intelligent machine" that oversees the generation of the adaptive spaces (Nadin, Novak 1987).

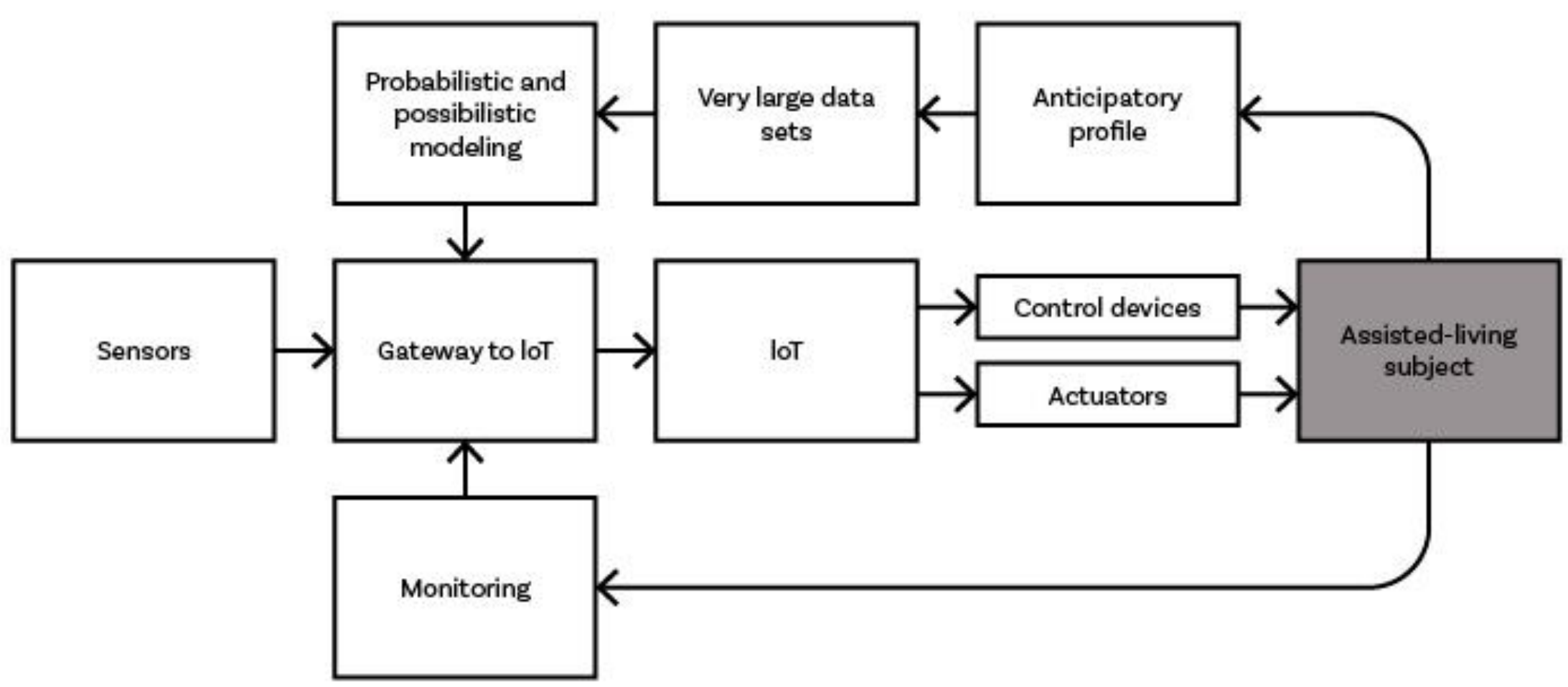

Figure 9. Data processing architecture integrates ML, IoT, probabilistic and possibilistic modelling

The "intelligence of the room" is expressed in its capability to identify a mood, a need, a desire, or an intended activity (cf. Figure 9). Data processing architecture integrates ML, IoT, probabilistic and possibilistic modelling. The integrated system learns to update and decide autonomously in order to meet the occupant's demands and expectations. Machine Learning will act as an interface for all data (Naz 2021). The learning process is both cognitive and perception-based. Recent advances in Deep Learning-based ambient assisted living (Monekosso et al. 2015, Abrantes et al. 2021) show that even chronic conditions (e.g., cardiovascular disease) can be ameliorated by providing adaptive spaces. This is a direction to be pursued.